\documentclass[prb,twocolumn,showpacs,floatfix,superscriptaddress]{revtex4}
\usepackage{graphics,epsfig,amsfonts,amssymb,amsmath,color}
\begin{document}
\title{Coupling of the $A_{1g}$ As-phonon to magnetism in iron pnictides}
\author{N.A. Garc\'ia-Mart\'inez}
\affiliation{Instituto de Ciencia de Materiales de Madrid,
ICMM-CSIC, Cantoblanco, E-28049 Madrid (Spain).}
\author{B. Valenzuela}
\affiliation{Instituto de Ciencia de Materiales de Madrid,
ICMM-CSIC, Cantoblanco, E-28049 Madrid (Spain).}
\author{S. Ciuchi} 
\affiliation{Dipartimento di Scienze Fisiche e Chimiche\\ 
Universit\`a dell'Aquila, CNISM and Istituto Sistemi Complessi CNR, 
via Vetoio, I-67010 Coppito-L'Aquila (Italy)} 
\affiliation{Istituto de Sistemi Complessi, U.O.S. Sapienza, CNR, v. dei
Taurini 19, 00185 Roma, Italy.}
\author{E. Cappelluti}
\affiliation{Istituto de Sistemi Complessi, U.O.S. Sapienza, CNR, v. dei
Taurini 19, 00185 Roma, Italy.}
\affiliation{Instituto de Ciencia de Materiales de Madrid,
ICMM-CSIC, Cantoblanco, E-28049 Madrid (Spain).} 
\author{M.J. Calder\'on}
\affiliation{Instituto de Ciencia de Materiales de Madrid,
ICMM-CSIC, Cantoblanco, E-28049 Madrid (Spain).}
\author{E. Bascones}
\affiliation{Instituto de Ciencia de Materiales de Madrid,
ICMM-CSIC, Cantoblanco, E-28049 Madrid (Spain).}

\date{\today}
\begin{abstract}
Charge, spin and lattice degrees of freedom are strongly entangled in iron
superconductors.
A neat consequence of this entanglement is the behavior of the $A_{1g}$ As-phonon resonance
in the different polarization symmetries of Raman spectroscopy when undergoing
the magneto-structural transition.
In this work we show that the observed behavior
could be a direct consequence of the coupling of the phonons with the electronic excitations
in the anisotropic magnetic state.
We discuss this scenario within a five orbital tight-binding model coupled to
phonons via the  dependence of the Slater-Koster parameters on the As
position.
We identify two qualitatively different channels of the
electron-phonon interaction: a geometrical one related to the
Fe-As-Fe angle $\alpha$ and another one associated with the modification upon As displacement
of the Fe-As energy integrals $pd\sigma$ and $pd\pi$.
While both mechanisms result in a finite $B_{1g}$
response,  the behavior of the phonon intensity in the
$A_{1g}$ and $B_{1g}$ Raman polarization geometries is
qualitatively different when the coupling is driven by the angle or by the
energy integral dependence. We discuss our results in view of the experimental reports.
\end{abstract}
\maketitle

\section{Introduction}

There is ample experimental evidence that iron pnictides present a rich interplay between
charge, lattice and magnetic degrees of freedom. 
The magnetic transition is commonly accompanied
by a structural one. Several phononic spectroscopic signatures
show unconventional behavior in the magnetic
state.~\cite{canfieldprb08,letaconprb09,chauviere09,akrapprb09,zhang10,basovprb11,uchida2011,gallaisprb11}
At the theoretical level,
ab-initio calculations show that the lattice
constants and the phonon frequencies depend sensitively on the presence of
magnetism,~\cite{yildirimphysC09,boeriprb10} and the comparison with experimental measurements is improved
when magnetism is included in the
calculations.~\cite{zbiriprb09,reznikprb09,hahn09,yildirimphysC09,mittalprb13,hahn13} 
Furthermore, the electron-phonon
coupling has been shown to be enhanced by
magnetism.~\cite{yndurainprb09,zbiriprb09,zbiriJPCM10,huangprb10,
liapl11,boeriprb10,lijap12}
Within this scenario, the role of the spin degree of freedom in the
electron-phonon coupling and its possible relevance in the mechanism of
superconductivity has been emphasized in several works.~\cite{yndurainprb09,huangprb10,egamiAdvCondMatt10,gadermaier12}

The electronic and magnetic properties are especially sensitive to the
height of the pnictogen atom, which affects the band structure at the Fermi
level,~\cite{vildosola08,nosotrasprb09} the magnetic
moment,~\cite{yin08,yndurainprb09,yndurainepl11,yildirimphysC09,cruzprl10,
egamiAdvCondMatt10} and possibly the superconducting critical temperature and gap.~\cite{iyo08,zhao08,kuroki09-2,garbarino2011}
Accordingly,
the A$_{1g}$ As-phonon, which involves vibrations of the As atoms along
the c-axis (see Fig.~\ref{fig:red}) seems to play a special role. Coupling to
this phonon has been detected by ultrafast techniques.~\cite{kimultrafast2012,kumar12,avigo13,rettig13}  A rapid
development of the magnetic ordering upon the vibrational displacement of the
$A_{1g}$ As-phonon has been observed.~\cite{kimultrafast2012} Features in the
ARPES spectrum of 11 compounds with an energy scale close to the one of this
phonon have been interpreted in terms of polaron formation.~\cite{shenprl13}

Raman response represents a powerful tool for investigating
the properties of lattice dynamics.~\cite{hadjiev08,litvinchuk08,canfieldprb08,gallais08,zhang09,rahlenbeckprb09,chauviere09,zhang10,gallaisprb11,sugai12,revramanpnic12}
A significant narrowing of the $A_{1g}$ As-phonon linewidth at the
onset of magnetism has been reported, whereas both
softening and hardening of the phonon frequency with decreasing temperature have
been observed.~\cite{canfieldprb08,gallais08,rahlenbeckprb09}
Crucial information is also encoded in the intensity of
the phonon resonances.
In the undistorted paramagnetic state, the $A_{1g}$ As-phonon is active neither in the $B_{1g}$ nor in the  $B_{2g}$ polarization symmetries. When undergoing the magneto-structural transition, a strong phonon signal emerges in the $B_{1g}$ Raman response but not in $B_{2g}$.~\cite{gallaisprb11,sugai12} In 122 compounds  the $A_{1g}$ intensity shows a strong enhancement in the magnetic state.~\cite{canfieldprb08,chauviere09,gallaisprb11,sugai12} On spite of this, in BaFe$_2$As$_2$ the $B_{1g}$ intensity is about $1.5$ times larger than the $A_{1g}$ intensity.~\cite{gallaisprb11,sugai12} This is not accounted for by the orthorhombic distortion alone, in agreement with the small $B_{1g}$ Raman intensity observed below the non-magnetic structural transition in FeSe.~\cite{gnezdilov13}  

The aim of this paper is to analyze the unusual Raman response and the changes on the phonon properties in the magnetic state. 
We show that the dynamical electron-phonon coupling
can be responsible for large and unconventional anomalies
in the phonon Raman spectrum.
We focus on the out-of-plane As lattice vibrations (the A$_{1g}$ As-phonon)
for which we explicitly calculate the electron-phonon coupling
within the context of a tight-binding Slater-Koster formalism.
The electron-phonon coupling
is formally split into two main qualitatively different
contributions: 
$i$) a purely geometrical one ($\hat{g}^\alpha$), related to the variation
of the Fe-As-Fe angle $\alpha$; and
$ii$) a second one ($\hat{g}^{pd}$)
coming from the variation of the
Slater-Koster energy integrals $pd\sigma$ and
$pd\pi$ (see Ref.~[\onlinecite{nosotrasprb09}]) 
upon the modulation of the Fe-As distance. We consider the $(\pi,0)$ magnetic state with magnetic moments ordered antiferromagnetically in the $x$ direction and ferromagnetically in the $y$ direction. 
Magnetism is included at the  mean-field
Hartree-Fock level,~\cite{nosotrasprl10,nosotrasprb12-2} and we study
separately  its interplay with the sources ($i$)-($ii$) of the
electron-phonon interaction.
The Raman response is evaluated in the paramagnetic
and in the $(\pi,0)$ magnetic states using the proper generalization
of the charge-phonon theory \cite{rice-chphonon-prl76} discussed
in Refs.~[\onlinecite{emm,emmprb12}].

Under generic conditions and excluding any static lattice distortion,
the coupling of the phonons
with the electronic excitations in the magnetic phase is able
by itself to induce a
Raman intensity in the $B_{1g}$ Raman polarization. This is due to the symmetry breaking
in the anisotropic magnetic state. Only the electron-phonon coupling $\hat g^\alpha$ accounts for the fact that the $B_{1g}$ Raman signal can be larger than the $A_{1g}$ Raman signal, the latter strongly decreasing in the magnetic state. A large enhancement in $A_{1g}$ with magnetism appears when coupling electrons and phonons via $\hat{g}^{pd}$.
Based on symmetry considerations, we argue that in the double stripe
magnetic order of FeTe the out-of-plane $A_{1g}$ vibrations of the Te atoms
will show finite Raman intensity only within the $B_{2g}$ polarization set-up
and not in the $B_{1g}$ one.
With increasing interactions the phonon frequency softens when entering into the magnetic state  but hardening is observed for the $\hat{g}^{pd}$ coupling for the largest values of the on-site electronic interactions $U$ considered.

\section{Theory}
\subsection{Model}
\label{sec:model}
The Hamiltonian we use to study the electron-phonon coupling on the Fe superconductors has three terms
\begin{equation}
H=H_0+H_{\rm ph} +H_{U}.
\label{eq:hamiltonian}
\end{equation}
$H_0$ is the five Fe d-orbital tight-binding model for the Fe-As planes,
obtained after eliminating the As degree of freedom and 
previously proposed in Ref.~[\onlinecite{nosotrasprb09}].  $H_{\rm ph}$ is the phonon
part including the free phonon and the electron-phonon interaction, and $H_U$
contains the electronic interactions. The one Fe unit cell is used with $x$ and
$y$ directions along the nearest neighbor Fe-Fe bonds. In the following we describe in detail each of the terms of the Hamiltonian. 
\begin{figure}
\leavevmode
\includegraphics[clip,width=0.3\textwidth]{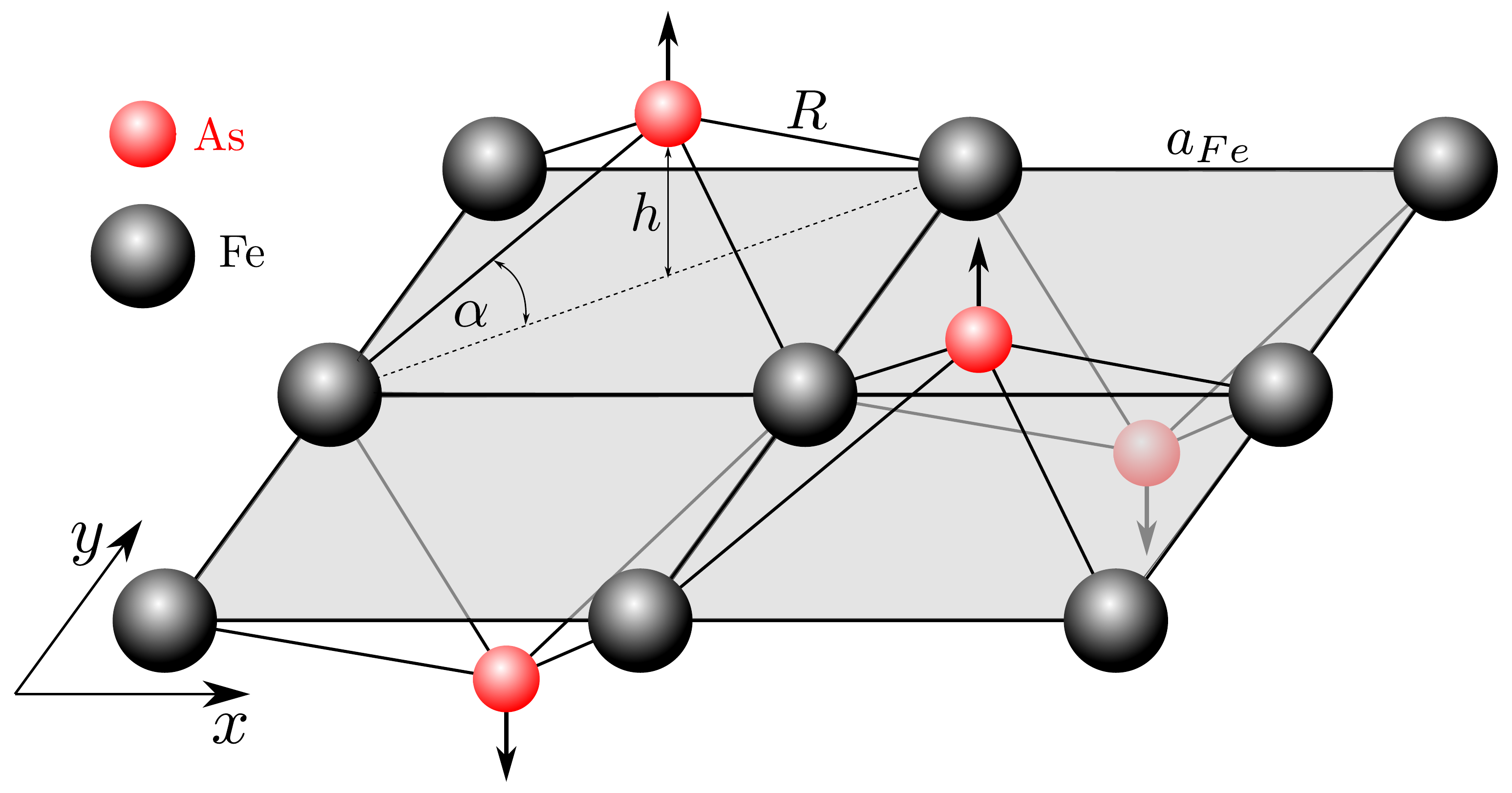}
\caption{ 
(Color online) Sketch of the lattice structure showing the $A_{1g}$ As-phonon in
iron arsenides. The Fe-Fe nearest neighbor distance $a_{Fe}$, the Fe-As-Fe angle
$\alpha$, the Fe-As distance $R$ and the As height with respect to the iron
plane $h$ are indicated.
 }
\label{fig:red} 
\end{figure}

\paragraph{Free electron part.} The band structure of the system is taken into account via the free electron term
$H_0$
\begin{equation}
H_0=\sum_{{\bf k},\mu,\nu,\sigma,r} F^r_{\mu\nu}({\bf k})  
t^r_{\mu\nu} c^{\dagger}_{{\bf k}\mu \sigma} c_{{\bf k}\nu\sigma}
+\sum_{i,\mu,\sigma}\epsilon_{\mu} c^{\dagger}_{i\mu\sigma} c_{i\mu\sigma} \, ,
\label{eq:H0}
\end{equation}
where $c^{\dagger}_{{\bf k}\mu \sigma}$ creates an electron on the Fe d-orbital $\mu$ with spin $\sigma$
and wave vector ${\bf k}$, and $c^{\dagger}_{i\mu\sigma}$ represents the same operator in real space.
$F^r_{\mu\nu}({\bf k})$ is the
electronic {\bf k}-dispersion relation\cite{nosotrasprb09} and $\epsilon_{\mu}$ is
the crystal field. 
$r$ labels the three different directions for the hoppings $t^r_{\mu\nu}$, with different {\bf k}-dispersions,
taken into account: between first neighbors in the x-direction, first neighbors
in the y-direction, and second neighbors. 

Direct Fe-Fe hoppings between their d-orbitals and indirect through the As
p-orbitals are considered. Indirect hopping is included to second order in
perturbation theory. \cite{nosotrasprb09} Within the Slater-Koster framework
considered,~\cite{slater54} the hopping parameters have explicit information on
the geometry of the pnictogen tetrahedra. These parameters depend on $\alpha$,
the angle between the Fe-As bond and the Fe plane (see Fig.~\ref{fig:red}) and
on the energy integrals.  The direct Fe-Fe hoppings depend on the
energy integrals $dd\sigma$, $dd\pi$ and $dd\delta$ between the Fe d-orbitals  while the
indirect (through the As) Fe-Fe hoppings depend on $pd\sigma$ and
$pd\pi$ between the Fe d-orbitals and the As p-orbitals. These energy integrals are a function of
the relative distance between the constituent atoms. The analytic expressions
for all the hoppings $t^r_{\mu\nu}$ are given in Ref.~[\onlinecite{nosotrasprb09}].

The energy integrals and crystal
field $\epsilon_{\mu}$ parameters in Eq.~(\ref{eq:H0}) are chosen to correctly describe the bands,
their orbital compositions, the Fermi surface, and the modification induced by
$\alpha$ as provided by electronic structure calculations. Details are given in Ref.~[\onlinecite{nosotrasprb09}].
$H_0$ has {\it tetragonal} symmetry, i.e. the static orthorhombic distortion found in the
magnetic phase is not included in our calculations.

\paragraph{Phonon part.} $H_{\rm ph}$ is given by 
\begin{eqnarray}
H_{\rm ph}&=&\sum_{\bf q} \omega_{\bf q} a_{\bf q}^{\dagger} a_{\bf q} \nonumber\\
&+& \sum_{{\bf k,q},\mu,\nu, \sigma, M} g_{\mu\nu}^M ({\bf k,q}) c^{\dagger}_{{\bf k}+{\bf q}\mu \sigma} c_{{\bf k}\nu \sigma} \left( a_{\bf q}+a_{-\bf q}^{\dagger} \right),
\end{eqnarray}
where $a_{\bf q}^{\dagger}$ creates a phonon in the $A_{1g}$ As-mode at wave vector ${\bf q}$,
$\omega_{\bf q}$ is the phonon frequency  and
$g_{\mu\nu}^M({\bf k,q})$ the electron-phonon matrix element between orbitals
$\mu$ and $\nu$. $M$ labels the type of interaction considered. The vertical displacement of the As atoms $\delta h$ (squeezing and
elongating the tetrahedra) gives rise to a modification of $\alpha$ around the
equilibrium position $\alpha_0$ and to a variation of the energy integrals
$pd\sigma$ and $pd\pi$ (only the indirect hoppings are affected by this phonon).
The two interaction terms that arise are labelled $\hat{g}^{\alpha}$ and 
$\hat{g}^{\rm pd}$, namely, $M=\alpha$ and $M=\rm pd$ respectively. 
In each of these cases, the electron-phonon interaction may arise from the
variation of the hopping $t^r_{\mu\nu}$ with phonon coordinates giving rise to {\it non-local} contributions, 
and from the variation of the crystal field $\epsilon_{\mu}$ resulting in {\it local} contributions: $g^M=g^{M,\rm loc}+g^{M,\rm non-loc}$.

The crystal field $\epsilon_{\mu}$ can be decomposed in a term which includes the electrostatic interactions between the ions in the system $\epsilon^{Coul}_\mu$ and a contribution that depends on the As orbital energies $\epsilon^{ind}_{\mu}$.  The dependence of $\epsilon^{Coul}_\mu$ on the As position is beyond the scope of the present study and is neglected here. We only consider the dependence of $\epsilon^{ind}_{\mu}$ (see Appendix~\ref{app:A}).

The non-local part $\hat{g}^{\alpha, \rm non-loc}$ involves the derivatives of the hoppings with respect to $\alpha$ [straightforwardly calculated from $h$, the distance between the As atoms and the Fe plane, and $a_{Fe}$, the Fe-Fe nearest neighbor distance, as $\alpha=\arctan (\sqrt 2 h/a_{Fe})$, see Fig.~\ref{fig:red}] and a form factor $\tilde F^r_{\mu\nu}({\bf k},{\bf q})$ which depends on the symmetry of the lattice and the orbitals. As we are here interested in the Raman response,
$\bf q=0$ and  $\tilde F^r_{\mu\nu}({\bf k},{\bf q})=F^r_{\mu\nu}({\bf k})$. In the orbital basis, the $\hat{g}^{\alpha, \rm non-loc}$ electron-phonon coupling is given by
\begin{equation}
g^{\alpha, \rm non-loc}_{\mu\nu}({\bf k})=\sum_r F^r_{\mu\nu}({\bf k}) \, {\partial t^r_{\mu\nu} \over {\partial \alpha}} \delta \alpha\, ,
\label{eq:galpha}
\end{equation}
with $\delta \alpha= {\partial \alpha \over {\partial h}}\delta h$ and form factors $F^r_{\mu\nu}({\bf k})$ as in Eq.~(\ref{eq:H0}).
Analogously 
\begin{equation}
g^{\alpha, \rm loc}_{\mu \mu}= {\partial \epsilon^{ind}_{\mu} \over {\partial \alpha}} \delta \alpha \, .
\label{eq:galpha_loc}
\end{equation}
This contribution to the interaction Hamiltonian appears as a constant (non {\bf k}-dependent) diagonal term. 

$\hat{g}^{\rm pd}$ involves the derivatives of the hoppings with respect to $pd\sigma$ and $pd\pi$. These integrals are a decaying function of $R$, the distance between Fe and As atoms. $R$ is related to $h$ as $R=h/\sin\alpha$. We assume the $R$ dependence is the same for both $pd\sigma$ and $pd\pi$: $pd\sigma=C_{pd\sigma} f(R)$ and $pd\pi=C_{pd\pi} f(R)$. If $R_0$ is the Fe-As distance corresponding to the equilibrium angle $\alpha_0$, $pd\sigma_0=C_{pd\sigma} f(R_0)$ and $pd\pi_0=C_{pd\pi} f(R_0)$. Expanding around this equilibrium value $pd\sigma=pd\sigma_0 \left[1+{1\over f(R_0)} {\partial f(R)\over \partial R} \delta R\right]$, and equivalently for $pd\pi$. Therefore, $\delta pd\sigma=pd\sigma_0 {1\over f(R_0)} {\partial f(R)\over \partial R}{\partial{R} \over{\partial h} }\delta h $ and $\delta pd\pi=pd\pi_0 {1\over f(R_0)} {\partial f(R)\over \partial R} {\partial{R} \over{\partial h} }\delta h$. Consequently, $\delta pd\pi={pd\pi_0 \over pd\sigma_0} \delta pd\sigma$. 
Using these relations, in the orbital basis
\begin{equation}
g^{pd,\rm non-loc}_{\mu\nu}({\bf k})=
\sum_r F^r_{\mu\nu}({\bf k})\delta pd\sigma \,  
\left({\partial t^r_{\mu\nu} \over {\partial pd\sigma}} + 
{\partial t^r_{\mu\nu} \over {\partial pd\pi}} \frac{ pd\pi_0}{ pd\sigma_0} \right)\, ,
\label{eq:gpd}
\end {equation}
\begin{equation}
g^{pd,\rm loc}_{\mu \mu}=\delta pd\sigma \,  \left({\partial \epsilon^{ind}_{\mu} \over {\partial pd\sigma}} +{\partial \epsilon^{ind}_{\mu} \over {\partial pd\pi}} \frac{ pd\pi_0}{ pd\sigma_0} \right) \, .
\label{eq:gpd_loc}
\end {equation}
\noindent

Here we take $f(R)=1/R^4$. This dependence is valid  assuming the $p$ and $d$
orbitals are very localized and they only couple through plane-wave corrections
to the atomic state wave-functions.~\cite{harrison-book} The pnictides
have a strong covalent character that may invalidate the localization
assumption, hence this particular functional dependence must be taken with
caution.  Therefore,
a direct quantitative comparison between $\hat{g}^{\alpha}$
and $\hat{g}^{\rm pd}$ would not be reliable, hence we
present the results for the two electron-phonon
interactions separately.

\paragraph{Correlation part.} 
$H_U$ includes the local interactions (intraorbital $U$, Hund's coupling $J_H$
and interorbital $U'=U-2J_H$) and is treated within Hartree-Fock mean field
approximation with focus on the $(\pi,0)$ antiferromagnetic
state (see Refs.~[\onlinecite{nosotrasprl10,nosotrasprb12-2}] for details).  The Hartree-Fock self-consistency includes the electronic degrees of freedom and not the phonons. The model without phonons ($H_0+H_U$) has been previously used to
study the magnetic phase diagram as a function of $U$ and $J_H/U$ within a
Hartree-Fock approximation.~\cite{nosotrasprl10,nosotrasprb12,nosotrasprb12-2}
With increasing $U$ a metallic AF $(\pi,0)$ state arises. For a narrow range of
values of $U$, the system can be described as itinerant, but a strong orbital
differentiation develops for larger values of $U$.~\cite{nosotrasprb12-2} In the
orbital differentiated region, the $3z^2-r^2$, $x^2-y^2$ and $zx$ orbitals are
itinerant while $xy$ and $yz$ are gapped at half-filling at the Fermi energy.
These results are consistent with experimental\cite{shen2012,sudayama2012} and
theoretical\cite{liebsch2010,aichhorn2010,yin11,si2012}  reports of different
renormalization values for the bands depending on their orbital character. In
our calculations,~\cite{nosotrasprb12-2} for $J_H/U=0.25$ the itinerant region
occurs for  $1.45\,$eV$<U<1.7\,$eV and the orbital differentiation for
$U>1.7\,$eV. 

\subsection{Phonon-mediated Raman scattering theory}
\label{sec:raman}
\begin{figure}
\leavevmode
\includegraphics[clip,width=0.4\textwidth]{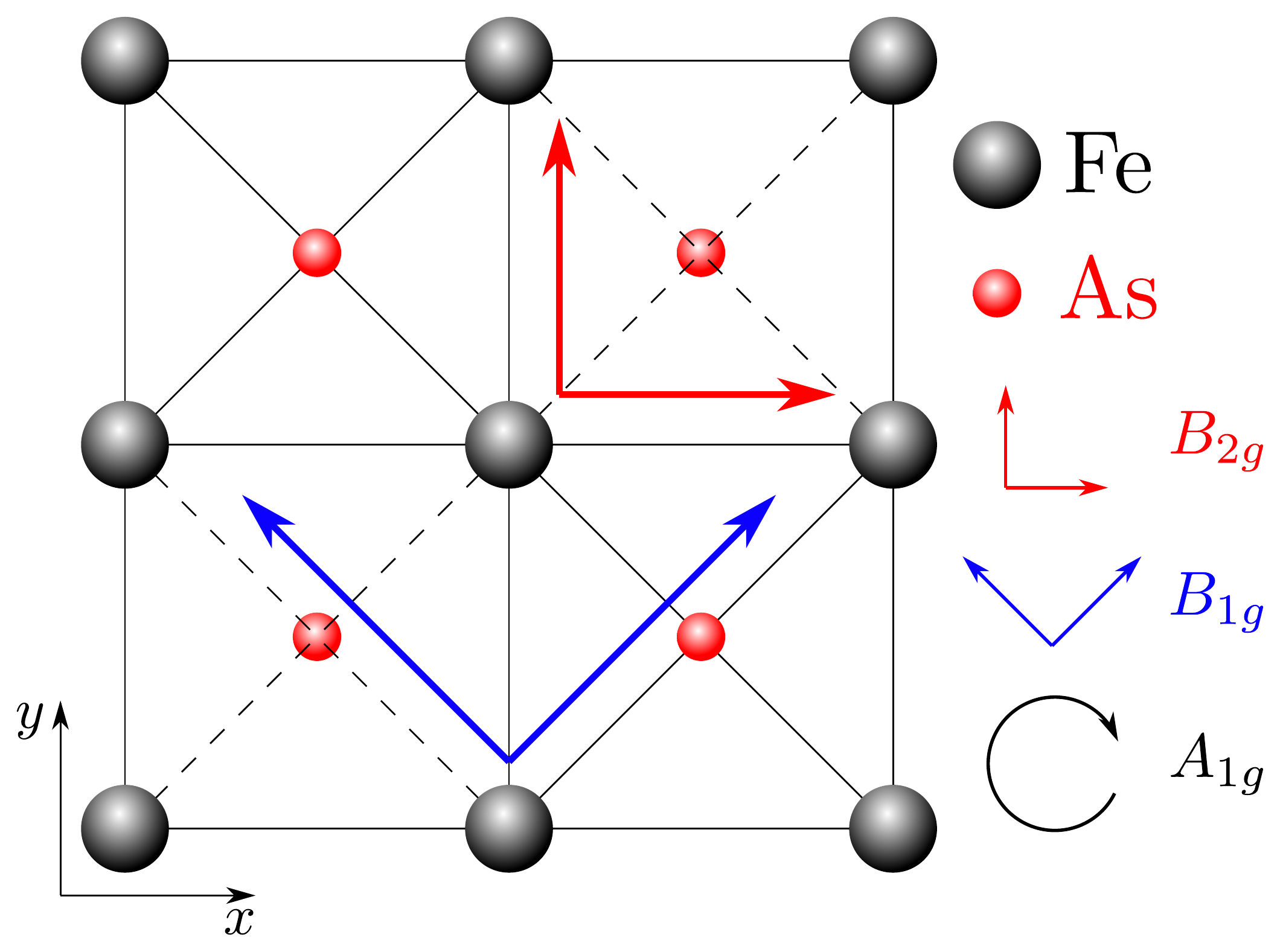}
\caption{(Color online) $A_{1g}$, $B_{1g}$ and $B_{2g}$ Raman symmetries for the Fe-As layer. We work in the one Fe unit cell with $x$ and $y$ directions along the nearest neighbor Fe-Fe bonds.}
\label{fig:b1gb2ga1g} 
\end{figure}

Raman scattering measures the total cross section of the inelastic scattering of electrons
\begin{equation}
\frac{\partial^2 \sigma}{\partial \Omega \partial \omega_S}=hr_0^2 \frac{\omega_S}{\omega_I}S(i\Omega \rightarrow \Omega+i0)
\end{equation}
with $\omega_I$ and $\omega_S$ the frequency of the incident and the scattered light respectively and $r_0$ the Thomson radius. The Raman intensity can be related to the imaginary part of the Raman response function
\begin{equation}
S^\lambda(\Omega)=-\pi^{-1} (1+n(\Omega,T)) \mathrm{Im} \chi^\lambda(\Omega)
\label{eq:raman}
\end{equation}
\noindent
with $\lambda=B_{1g},B_{2g},A_{1g}$ the symmetries of the squared lattice point
group depending on the incident and scattered photon polarizations represented
in Fig.~\ref{fig:b1gb2ga1g}. The symmetries are defined with the $x$ and $y$ axis
along the Fe-Fe nearest-neighbors, as in the Hamiltonian. This  definition is
different from the one used in some experimental papers.~\cite{gallaisprb11}

\begin{figure}
\leavevmode
\includegraphics[clip,width=0.47\textwidth]{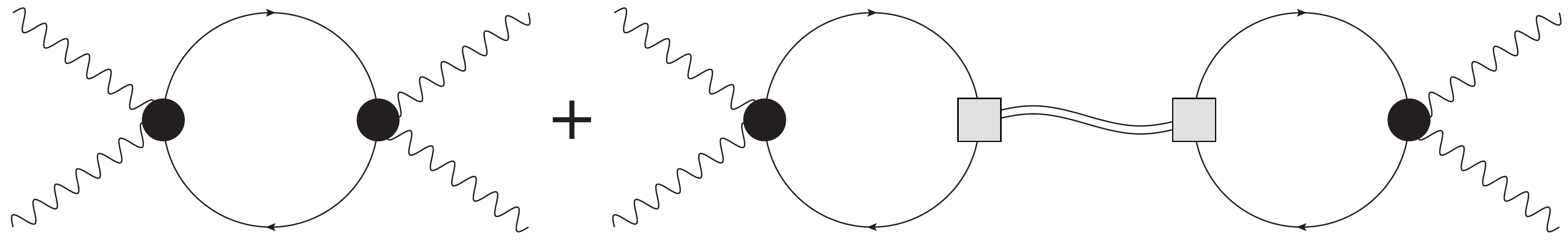}
\caption{Electronic Raman diagram (left) and phonon-mediated Raman diagram
(right). The wavy line is the photon propagator, the double wavy line is the phonon propagator.
The circle stands for the Raman vertex and the square for the
electron-phonon interaction, either $\hat g^\alpha$ or $\hat g^{pd}$. 
The electronic Raman diagram was studied in Ref.~[\onlinecite{nosotrasprb13}]. }
\label{fig:ramandiagram} 
\end{figure}
Only non-resonant diagrams are included in the calculation of the Raman
response. To study the phonon contribution we use  the
charge-phonon theory originally proposed by M.J. Rice\cite{rice-chphonon-prl76}
for the optical conductivity and recently used\cite{emm,emmprb12} to study the Raman response in graphene. The Raman response includes the diagrams shown in
Fig.~\ref{fig:ramandiagram}: the electronic bubble contribution
$\chi^\lambda_{el-el}$ (left) and the charge-phonon diagram $\Delta
\chi^\lambda_{ph}$ (right),
\begin{equation}
\chi^\lambda (\Omega)=\chi^{\lambda}_{el-el}(\Omega)+\Delta \chi^\lambda_{ph}(\Omega) \,.
\label{eq:chilambda}
\end{equation}

The electronic Raman response $\chi^\lambda_{el-el}$ was studied in Ref.~[\onlinecite{nosotrasprb13}] and is not discussed in this work. Here we concentrate our attention on the 
phonon-mediated Raman response that can be expressed as a sum on all electron-phonon channels $M,M^{\prime}$:
$\Delta \chi^\lambda_{ph}(\Omega)=\sum_{M,M^\prime} \Delta \chi^\lambda_{ph,M,M^{\prime}}$,

\begin{equation}
\Delta \chi^\lambda_{ph,M,M'} (\Omega)=\chi^{\lambda}_M (\Omega) D_0(\Omega) \chi^{*\lambda}_{M'} (\Omega) \, ,
\end{equation}

\noindent
with $D_0(\Omega)= D_0({\bf q}=0,\Omega)$ the phonon propagator.
To address separately the contribution of a single electron-phonon channel
we focus on the diagonal part of $\Delta \chi^\lambda_{ph,M,M^{\prime}}=\Delta \chi^\lambda_{ph,M}$. 

In the vicinity of the resonance we can approximate
\begin{equation}
D_0(\Omega)=\frac{1}{(\Omega-\Omega_0)+i\Gamma_0} \,.
\end{equation}
with $\Omega_0$ the phonon frequency and $\Gamma_0$ the phonon scattering rate.

$\chi^\lambda_M$ is a mixed response which includes both the electron-phonon couplings $\hat{g}^M$ defined in Eqs.~(\ref{eq:galpha})-(\ref{eq:galpha_loc}) and (\ref{eq:gpd})-(\ref{eq:gpd_loc}) and the Raman vertex  $\gamma^\lambda$. The $\gamma^\lambda$ vertices encode information of the incident and scattered light, and the point group symmetry of the squared lattice.~\cite{devereauxreview}  In the orbital basis they are given by:\cite{nosotrasprb13} 
\begin{eqnarray}
& \gamma_{\mu\nu}^{B_{1g}}(\bf {k}) &=\frac{\partial^2\epsilon_{\mu\nu}({\bf k})}{\partial k^2_x}-\frac{\partial^2\epsilon_{\mu\nu}({\bf k})}{\partial k^2_y} \, ,
\label{eq:b1g}\\ 
& \gamma_{\mu\nu}^{B_{2g}}(\bf {k}) &= \frac{\partial^2\epsilon_{\mu\nu}({\bf k})}{\partial k_x \partial k_y} \, ,
\label{eq:b2g}\\ 
& \gamma_{\mu\nu}^{A_{1g}}(\bf {k}) &= \frac{\partial^2\epsilon_{\mu\nu}({\bf k})}{\partial k^2_x}+\frac{\partial^2\epsilon_{\mu\nu}({\bf k})}{\partial k^2_y} \, .
\label{eq:a1g}
\end{eqnarray}

Separating real and imaginary parts $\chi^\lambda_M=\chi'^\lambda_M+i\chi''^\lambda_M$
\begin{eqnarray}
\chi'^\lambda_M(\Omega)&=&\frac{1}{V}\sum_{{\bf k}\sigma nn'}\gamma^\lambda_{nn'}({\bf k}) g^{M*}_{nn'}({\bf k})  
\left(f(E_n({\bf k}))-  f(E_{n'}({\bf k}))\right)\nonumber\\ 
&\times&(\frac{\Omega+E_n({\bf k})-E_{n'}({\bf k})}{(E_n({\bf k})-E_{n'}({\bf k})+\Omega)^2+\eta^2}
\nonumber \\
&+&\frac{-\Omega+E_n({\bf k})-E_{n'}({\bf k})}{(E_n({\bf k})-E_{n'}({\bf k})-\Omega)^2+\eta^2})
\label{eq:rechi}
\end{eqnarray}
\begin{eqnarray}
&\chi''^\lambda_M&(\Omega)=-\frac{\pi}{V}\sum_{{\bf k}\sigma nn'}\gamma^\lambda_{nn'}({\bf k}) g^{M*}_{nn'}({\bf k}) 
\left(f(E_n({\bf k})) - f(E_{n'}({\bf k}))\right)
\nonumber
\\
&\times&\left(\delta(\Omega + E_n({\bf k})-E_{n'}({\bf k}))
- \delta(-\Omega + E_n({\bf k})-E_{n'}({\bf k}))\right )\nonumber\\
\label{eq:imchi} 
\end{eqnarray}
\noindent
with $\delta$ functions broadened by $\eta$. Here $V$ is the volume, $E_n$,
$E_{n'}$ label the energies of the bands  $n$ and $n'$, $f(E)$ is the Fermi
function, $g^M_{nn'}({\bf k})=\sum_{\mu\nu}a^*_{\mu n} ({\bf
k})g^M_{\mu\nu}({\bf k})a_{\nu n'}({\bf k})$ and $\gamma^\lambda_{nn'}({\bf
k})=\sum_{\mu\nu}a^*_{\mu n}({\bf k}) \gamma^\lambda_{\mu\nu}({\bf k})a_{\nu
n'}({\bf k})$ with $a_{\mu n}$ the matrix which rotates between the orbital and
the band basis.~\cite{nosotrasprb13} The matrix elements $g^M_{nn'}({\bf k})$ and
$\gamma^\lambda_{nn'}({\bf k})$ determine whether the phonon is Raman active.

 $\Delta \chi^\lambda_{ph, M} (\Omega)$ can be rewriten as a function of the phonon intensity $I^\lambda_M$ and the Fano factor $q^\lambda_M$~[\onlinecite{emmprb12}]
    \begin{equation}
{\rm Im} \Delta \chi_{ph,M}^{\lambda} (\Omega)=-I^\lambda_M \frac{(q_M^{\lambda})^2-1+2(\frac{\Omega-\Omega_0}{\Gamma_0})q_M^{\lambda}}{(q_M^{\lambda})^2(1+(\frac{\Omega-\Omega_0}{\Gamma_0})^2)}
\label{eq:phonon}
\end{equation}
\noindent
with the intensity prefactor $I^\lambda_M$ and the Fano factor $q^\lambda_M$ given by
\begin{equation}
I^\lambda_M=\frac{\big(\chi'^{\lambda}_M(\Omega_0)\big)^2}{\Gamma_0} \, ,
\label{eq:intensity}
\end{equation}
\begin{equation}
q_M^\lambda=-\frac{\chi'^{\lambda}_M(\Omega_0)}{\chi''^{\lambda}_M (\Omega_0)} \, .
\label{eq:fano}
\end{equation}
\noindent
This formula will be used below to study the symmetry dependence and intensity of the $A_{1g}$ As-phonon.
If $|q_M^\lambda|$ is large, the phonon signal acquires a symmetric shape. If in Eq.~(\ref{eq:phonon}) we replace $\Omega=\Omega_0$, for $q_M^\lambda \gg 1$ we find ${\rm Im}\Delta \chi_{ph,M}^\lambda (\Omega_0)=-I_M^\lambda$. 
The Raman signal turns out positive when replaced in Eq.~(\ref{eq:raman}).

\subsection{Phonon self-energy}
\label{sec:phononselfenergy}
\begin{figure}
\leavevmode
\includegraphics[clip,width=0.2\textwidth]{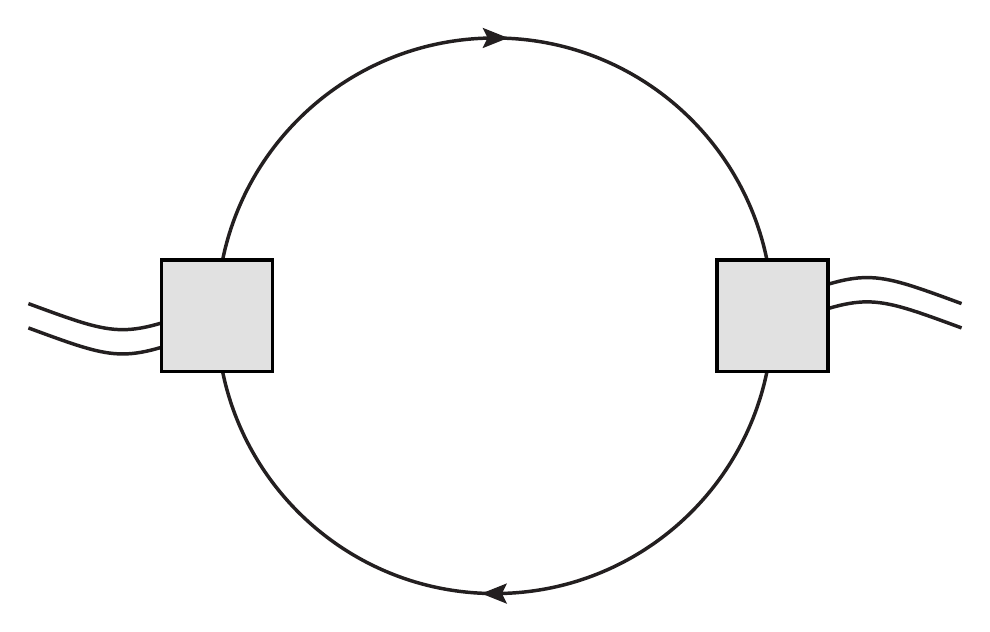}
\caption{Phonon self energy. The square stands for the electron phonon
interaction: either $\hat{g}^\alpha$ or $\hat{g}^{pd}$. 
}
\label{fig:phononselfenergy} 
\end{figure}
We study the ${\bf q}=0$ phonon self-energy contribution $\Pi_{M} (\Omega)=\Pi'_M(\Omega)+i\Pi''_M(\Omega)$ arising from the 
coupling to the electrons. The real part produces a hardening or softening of the phonon and the imaginary part contributes to the phonon broadening.
In the second order perturbation theory approximation,~\cite{mahan} (see Fig.~\ref{fig:phononselfenergy}) the phonon self-energy can be expressed as a sum 
on all the electron-phonon channels $M,M^{\prime}$:
$\Pi(\Omega)=\sum_{M,M^\prime} \Pi^\lambda_{ph,M,M^{\prime}}(\Omega)$.
We consider only the diagonal part of the self-energy $\Pi_M(\Omega)$ for which the real and imaginary part read
\begin{eqnarray}
\Pi'_M(\Omega)&=&\frac{1}{V} \sum_{{\bf k} \sigma nn'}|g^M_{nn'}({\bf k})|^2  \left(f(E_n({\bf k}))-f(E_{n'}({\bf k}))\right)
\nonumber\\
&\times&(\frac{\Omega+E_n({\bf k})-E_{n'}({\bf k})}{(E_n({\bf k})-E_{n'}({\bf k})+\Omega)^2+\eta^2}
\nonumber \\
&+&\frac{-\Omega+E_n({\bf k})-E_{n'}({\bf k})}{(E_n({\bf k})-E_{n'}({\bf k})-\Omega)^2+\eta^2})
\label{eq:rePi}
\end{eqnarray}
\noindent 

\begin{eqnarray}
\Pi''_M(\Omega)=-\frac{\pi}{V} \sum_{{\bf k}\sigma nn'}|g^M_{nn'}({\bf k})|^2 (f(E_n({\bf k}))-f(E_{n'}({\bf k}))\nonumber\\
\times \left ( \delta(\Omega + E_n({\bf k})-E_{n'}({\bf k})) -  \delta(-\Omega + E_n({\bf k})-E_{n'}({\bf k})) \right ) \,.
\nonumber 
\\
\label{eq:imPi}
\end{eqnarray}
\noindent 
A small broadening $\eta$, which also enters in the $\delta$ functions, has been
introduced. 

\section{Results}
\label{sec:results}
We have calculated the phonon contribution to the Raman response and the
correction to the phonon self-energy induced by the electron-phonon coupling in
the paramagnetic and $(\pi,0)$ antiferromagnetic states at zero temperature. We
study the A$_{1g}$ As-phonon and  consider the two electron-phonon couplings
introduced in Section ~\ref{sec:model}, $\hat{g}^\alpha$ and $\hat{g}^{pd}$. We
choose generic interactions to describe the iron pnictides, $J_H=0.25U$ with $U$
ranging from the paramagnetic phase $U<1.45$ eV, through the itinerant magnetic
phase $1.45$ eV$<U<1.7$ eV to the orbital differentiated region\cite{nosotrasprb12-2}
$U>1.7$ eV. $\alpha_0=35.3^\circ$ and $n=6$, corresponding to a regular
tetrahedra and undoped pnictides, are considered unless otherwise stated.  We
take $\delta h=0.02$ $\rm \AA$, and
$\Omega_{0}=20$ meV and $\Gamma_{0}=1$ meV for the phonon frequency and scattering rate (values are similar to the experimental
ones~\cite{canfieldprb08,gallais08,rahlenbeckprb09}). Within the Hartree-Fock
approximation used here to include the local interactions, the renormalization
of the bands is not properly accounted for. Comparison of {\em ab-initio}
electronic structure calculations and ARPES measurements render a factor of $3$
for the renormalization of the bands. 
Therefore once the ground state has been obtained, the energy bands are divided
by $3$ to account for the renormalization observed in ARPES
experiments\cite{lu08} and not reproduced at the Hartree-Fock level.

\subsection{Raman response}
\begin{figure}
\leavevmode
\includegraphics[clip,width=0.23\textwidth]{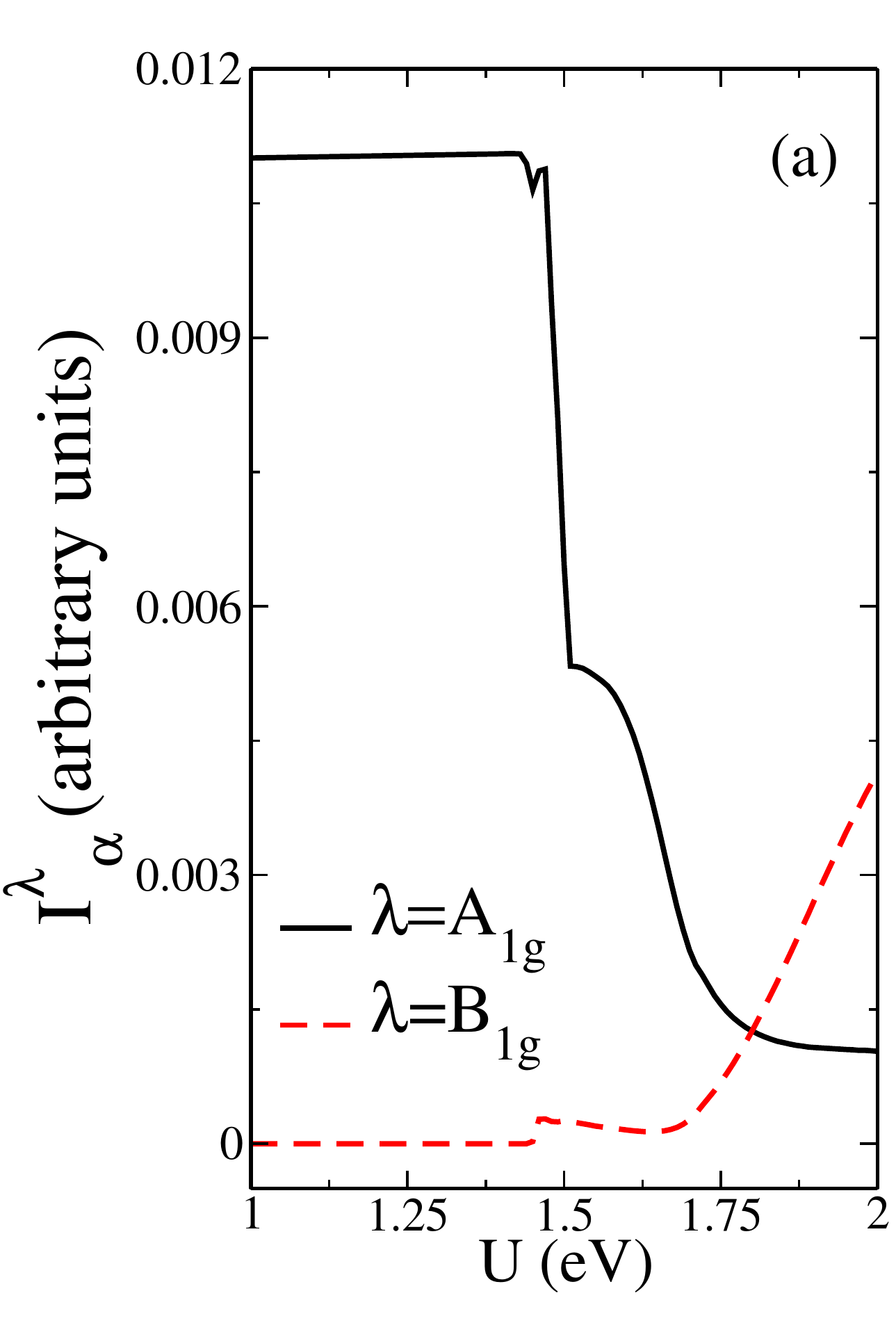}
\includegraphics[clip,width=0.23\textwidth]{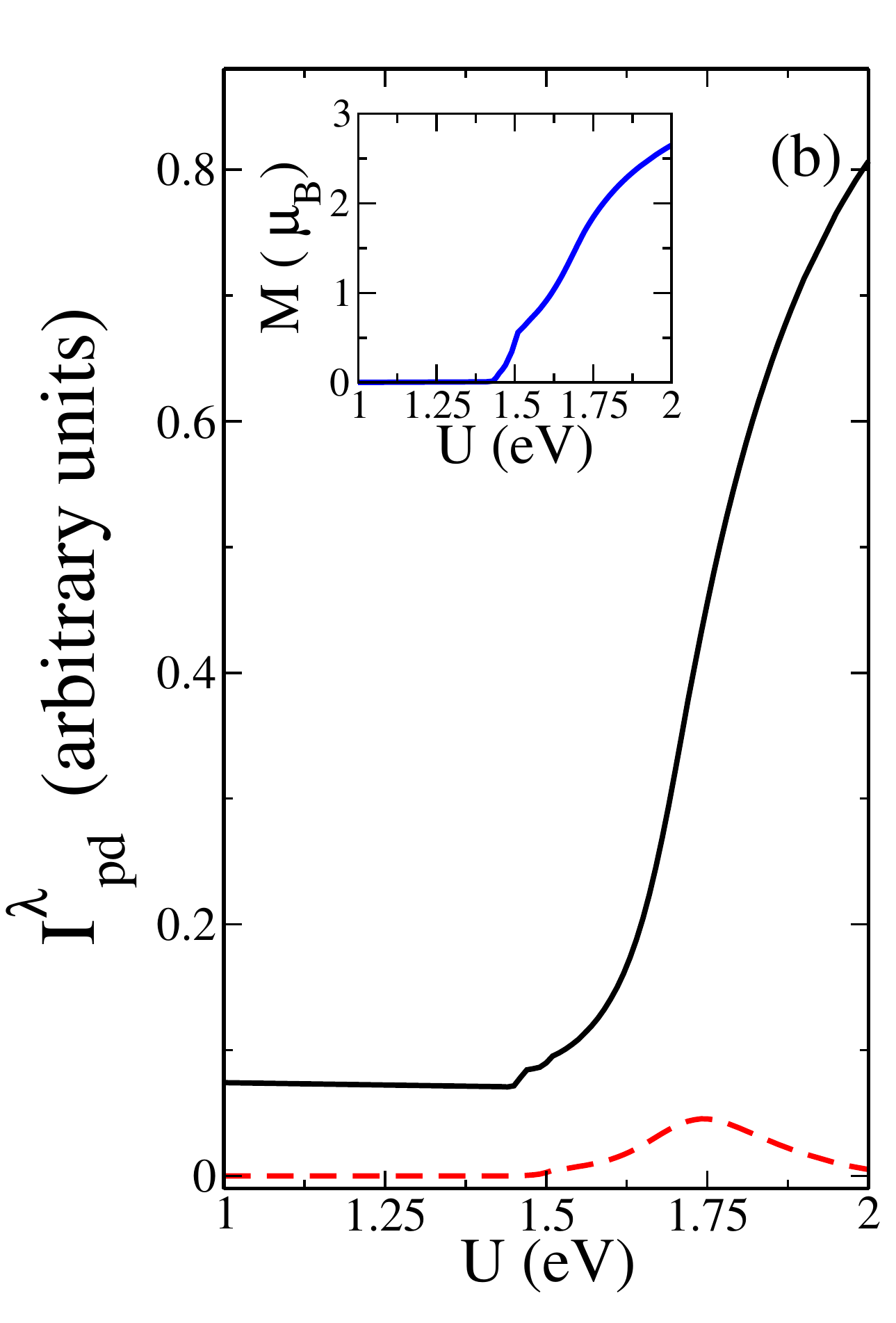}
\caption{(Color online) $A_{1g}$ and $B_{1g}$ phonon Raman intensities versus the on-site
interaction $U$  for the electron-phonon coupling $\hat{g}^\alpha$ (a) and 
$\hat{g}^{pd}$ (b).
Note that in the magnetic state, $\hat{g}^\alpha$
for $U > 1.8$ eV shows a bigger $B_{1g}$ response than the $A_{1g}$ one, while $\hat{g}^{pd}$ gives a strong enhancement of the $A_{1g}$ polarization signal.
Inset in (b) Magnetic moment as a function of the interaction $U$. $J_H/U=0.25$,
$\Omega_0=20$ meV, $\delta h=0.02 \,\rm\AA$,  $\Gamma_0=1$ meV, and $\eta=3$ meV
have been used. 
}
\label{fig:intensities} 
\end{figure}

Fig.~\ref{fig:intensities} is the main result of this work. It shows the
$A_{1g}$ and the $B_{1g}$ phonon Raman intensities due to the  couplings $\hat
g^\alpha$ (left) and $\hat{g}^{pd}$ (right). For both couplings $\hat g^ M$, 
the intensity $I^{A_{1g}}_M$ in the $A_{1g}$
polarization  is finite in both the paramagnetic and magnetic states while
$I^{B_{1g}}_M$ is finite only in the antiferromagnetic state. The $B_{2g}$
phonon intensity, not shown, vanishes in all the range of parameters. \cite{note-symmetry}
While $H_0$ is tetragonal, this symmetry is broken
in the anisotropic $(\pi,0)$ magnetic state. The $x$ and $y$ directions become
inequivalent due to the reorganization of the {\it electronic} degrees of
freedom. The $B_{1g}$ signal is antisymmetric under the $k_x \rightarrow k_y$
rotation, see Eq.~(\ref{eq:b1g}), and it is sensitive to $k_x$ being non
equivalent to $k_y$ in the magnetic state. This sensitivity results in a finite
$I^{B_{1g}}_M$. $B_{2g}$, however, is antisymmetric under either $k_x
\rightarrow -k_x$ or $k_y \rightarrow -k_y$, see Eq.~(\ref{eq:b2g}).
$I^{B_{2g}}_M$ is not sensitive to the breaking of the tetragonal symmetry in the $(\pi,0)$ state
and  remains zero. 

A strong change in the intensity is also observed in the $A_{1g}$ Raman polarization when entering in the magnetic state. The Raman intensity $I^{A_{1g}}_M$ is related to the real part of the mixed bubble at the phonon frequency $\chi'^{A_{1g}}_M (\Omega_0)$  via Eq.~(\ref{eq:intensity}). $\chi'^{A_{1g}}_M (\Omega_0)$, Kramers-Kr\"onig integral of $\chi''^{A_{1g}}_M (\Omega_0)$, is sensitive to the reorganization of the electronic structure in the magnetic state, especially close to the Fermi level, at energies comparable to $\Omega_0$.

The $A_{1g}$ and the $B_{1g}$ signals show a qualitatively different
behavior for the two different
electron-phonon couplings considered: $\hat{g}^\alpha$
[Fig.~\ref{fig:intensities}(a)] and $\hat{g}^{pd}$
[Fig.~\ref{fig:intensities}(b)]. In the magnetic state, $I^{A_{1g}}_{\alpha}$
decreases with respect to the intensity in the paramagnetic state, while
$I^{A_{1g}}_{pd}$ increases. $I^{B_{1g}}_{\alpha}$ increases in the magnetic
state while $I^{B_{1g}}_{pd}$ shows a bump as a function of $U$ getting close to
zero for $U=2$ eV.

\begin{figure}
\leavevmode
\includegraphics[clip,width=0.23\textwidth]{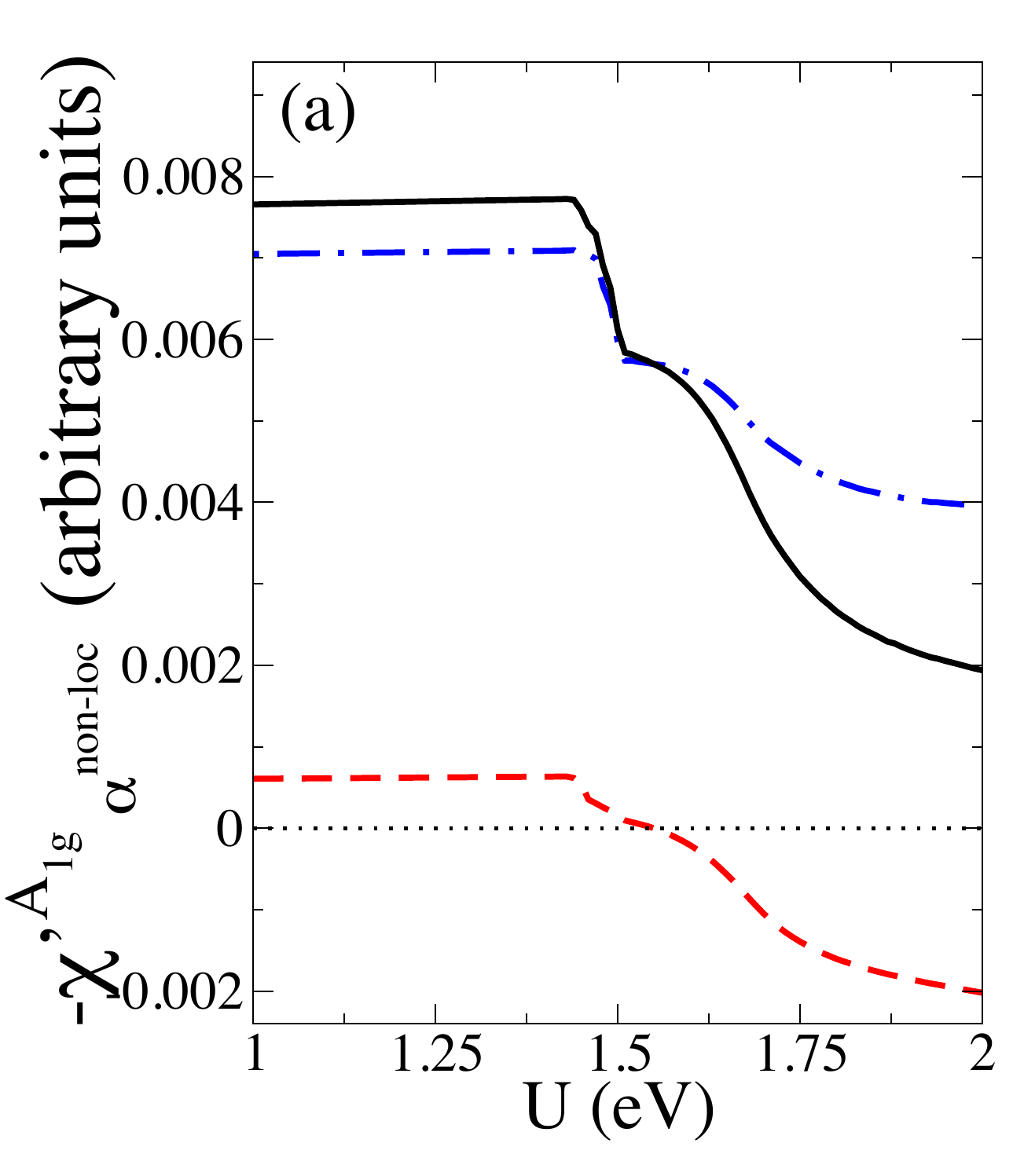}
\includegraphics[clip,width=0.23\textwidth]{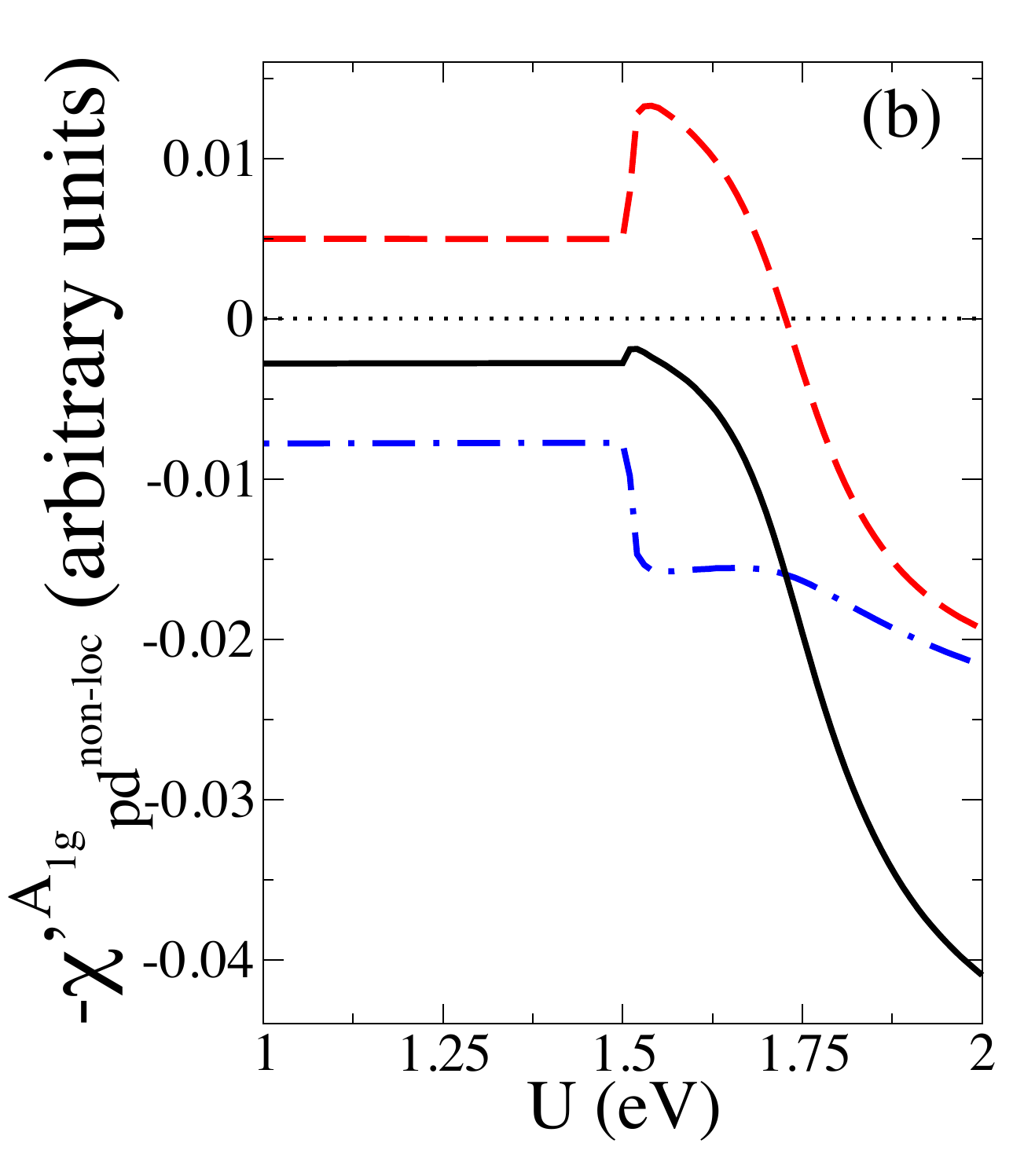}
\includegraphics[clip,width=0.23\textwidth]{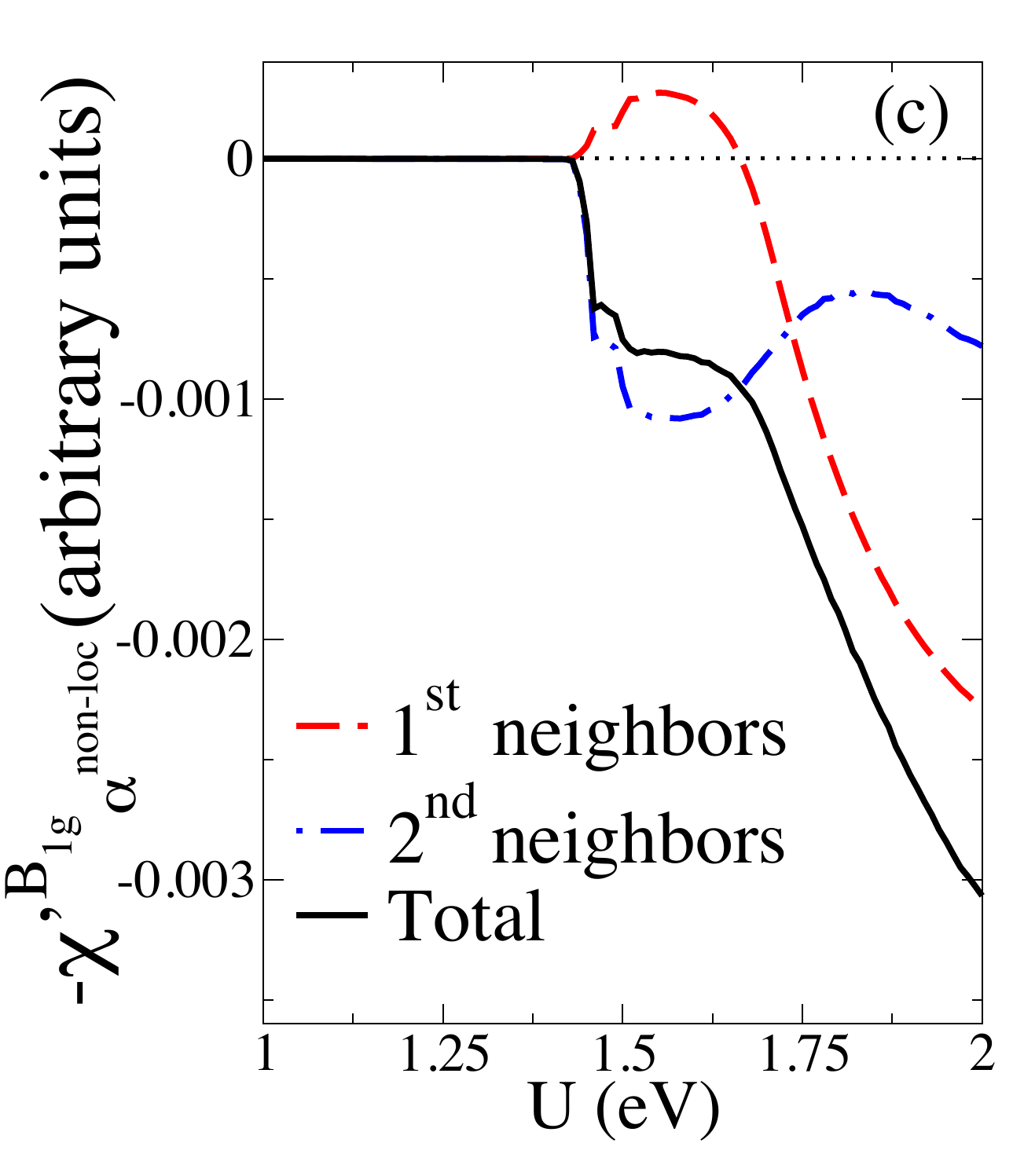}
\includegraphics[clip,width=0.23\textwidth]{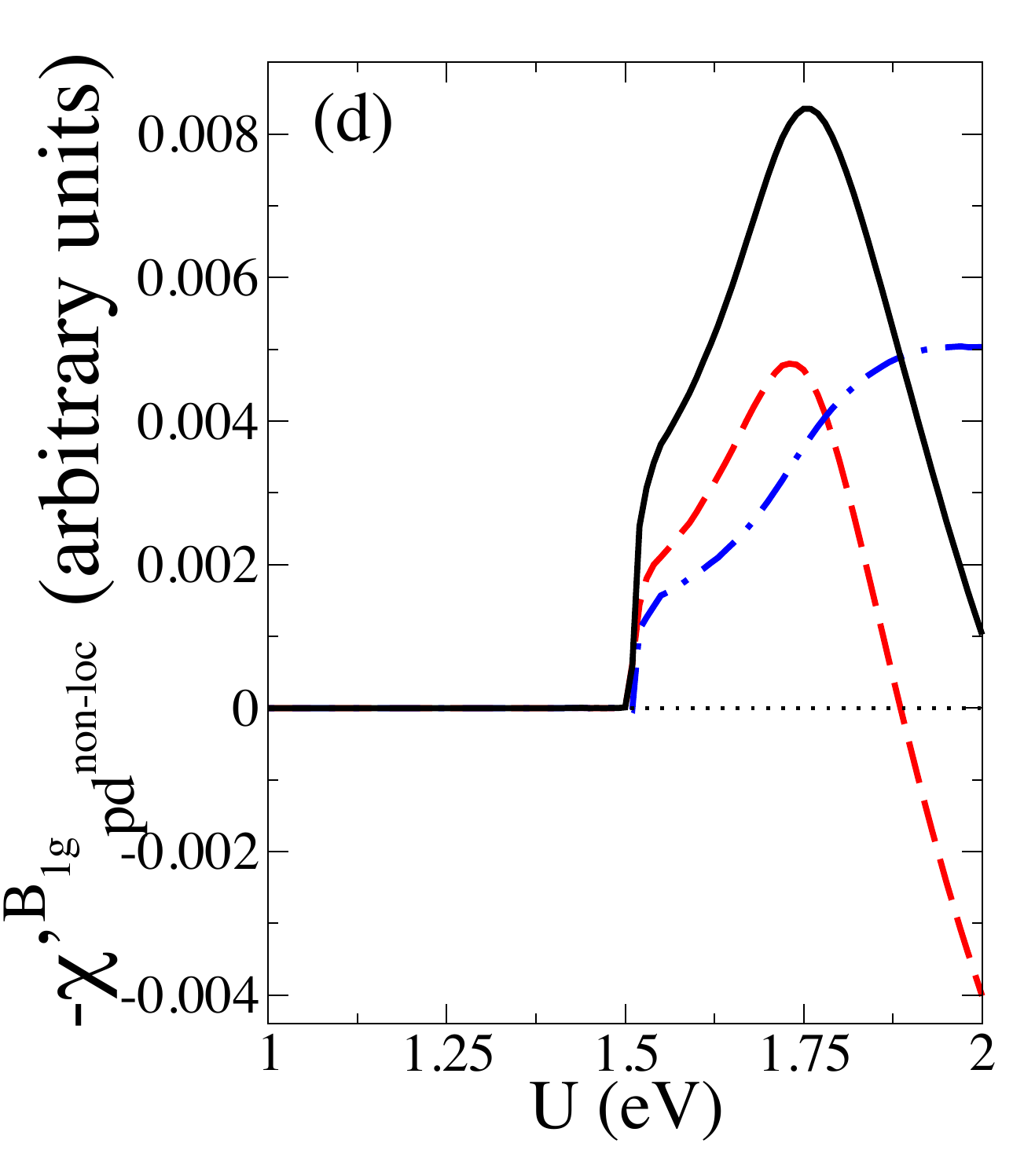}
\caption{(Color online) Real part of the non-local terms of the mixed bubble $\chi'^{\lambda}_M (\Omega_0)$: $\chi'^{A_{1g}}_{\alpha^{non-loc}}$ (a),    $\chi'^{A_{1g}}_{pd^{\rm non-loc}}$ (b), $\chi'^{B_{1g}} _{\alpha^{\rm non-loc}}$(c), $\chi'^{B_{1g}}_{pd^{\rm non-loc}}$ (d). These non-local terms appear to dominate the qualitative behavior of the Raman intensity shown in Fig.~\ref{fig:intensities}.  (a) First and second nearest neighbor contributions add up in the paramagnetic state but subtract in the magnetic state while in (b) they almost cancel in the paramagnetic state but add up in the magnetic state. As a result, since $I^{\lambda}_{M}$ is proportional to $\chi'^{\lambda}_M$ squared, $I^{A_{1g}}_{\alpha}$ decreases in the magnetic state while $I^{A_{1g}}_{pd}$ increases (Fig.~\ref{fig:intensities}). (d) In the magnetic state the contributions mostly add up but cancel at high values of $U$ for $\chi'^{B_{1g}}_{pd^{\rm non-loc}}$ which does not happen in (c), explaining the behavior of $I^{B_{1g}}_{M}$ with $U$. Same parameters as in Fig.~\ref{fig:intensities}.
 }
\label{fig:mixedbubble} 
\end{figure}

The non-local components of the electron-phonon interactions,
Eqs.~(\ref{eq:galpha}) and (\ref{eq:gpd}), appear to dominate the qualitative behaviour of the Raman intensity. This can be seen by comparing the total intensities in Fig.~\ref{fig:intensities} with the non-local terms of the mixed bubble $\chi'^{\lambda}_M (\Omega_0)$ in Fig.~\ref{fig:mixedbubble} ($I^{\lambda}_M$ and $\chi'^{\lambda}_M$ are related by Eq.~\ref{eq:intensity}). The relevant features shown in the solid lines of  Fig.~\ref{fig:mixedbubble} mimic the curves in Fig.~\ref{fig:intensities}. 

The non-local components $\chi'^{\lambda}_{M^{\rm non-loc}} (\Omega_0)$ result to be a linear combination of the same
${\bf k}$-dependent form factors $F^r_{\mu\nu}({\bf k})$
in both $M=\alpha$ and $M=pd$ phonon channels.
Therefore, for a given photon polarization, 
the difference in the behaviour of $\chi'^{\lambda}_{M}$ and $I^{\lambda}_M$ should be ascribed to the 
difference in the coefficients of the form factors.
In order to gain a deeper insight on this issue,
we have further decomposed
$\chi'^{\lambda}_{M^{\rm non-loc}} (\Omega_0)$ into first and second nearest
neighbor contributions. In  Fig.~\ref{fig:mixedbubble} we see
that the different behaviors observed in the Raman
intensity in Fig.~\ref{fig:intensities} for the two electron-phonon couplings as
a function of $U$ are partly a consequence of the fact that in some cases
different contributions add up and in other cases subtract.

The Raman intensities are not just a simple function of the
magnetic moment (see inset in Fig.~\ref{fig:intensities}(b)). Non-monotonic dependences in momentum are frequently found,
especially in the itinerant region. This becomes also clear when comparing the
spectrum corresponding to different angles $\alpha_0$, electron filling $n$ and
interactions (not shown). A change in the electron filling and Fe-As-Fe angle induces
changes in the band structure and in the transitions at energies close to
$\Omega_0$ and consequently in the Raman spectrum.
 
The Fano factors $q^\lambda_M$, not shown, corresponding to $\hat{g}^\alpha$ and
$\hat{g}^{pd}$ are calculated using the expression in Eq.~\ref{eq:fano}. For
$\hat{g}^\alpha$ and for both polarizations $B_{1g}$ and $A_{1g}$,  the Fano
factor is generally large and negative with values between $-40$ and $-30$ for
$U \geq 1.8$ eV. For $\hat{g}^{pd}$ the Fano factor is even larger and still
negative reaching around $-40$ just for $1.8 \le U \le 1.9$ eV in the $B_{1g}$
polarization. This Fano factor corresponds to an almost symmetric Lorentzian
form of the Raman phonon peak. For smaller values of $U$, $q^\lambda_M$ is
strongly dependent on the parameters. 

\subsection{Phonon self-energy}
\begin{figure}
\leavevmode
\includegraphics[clip,width=0.23\textwidth]{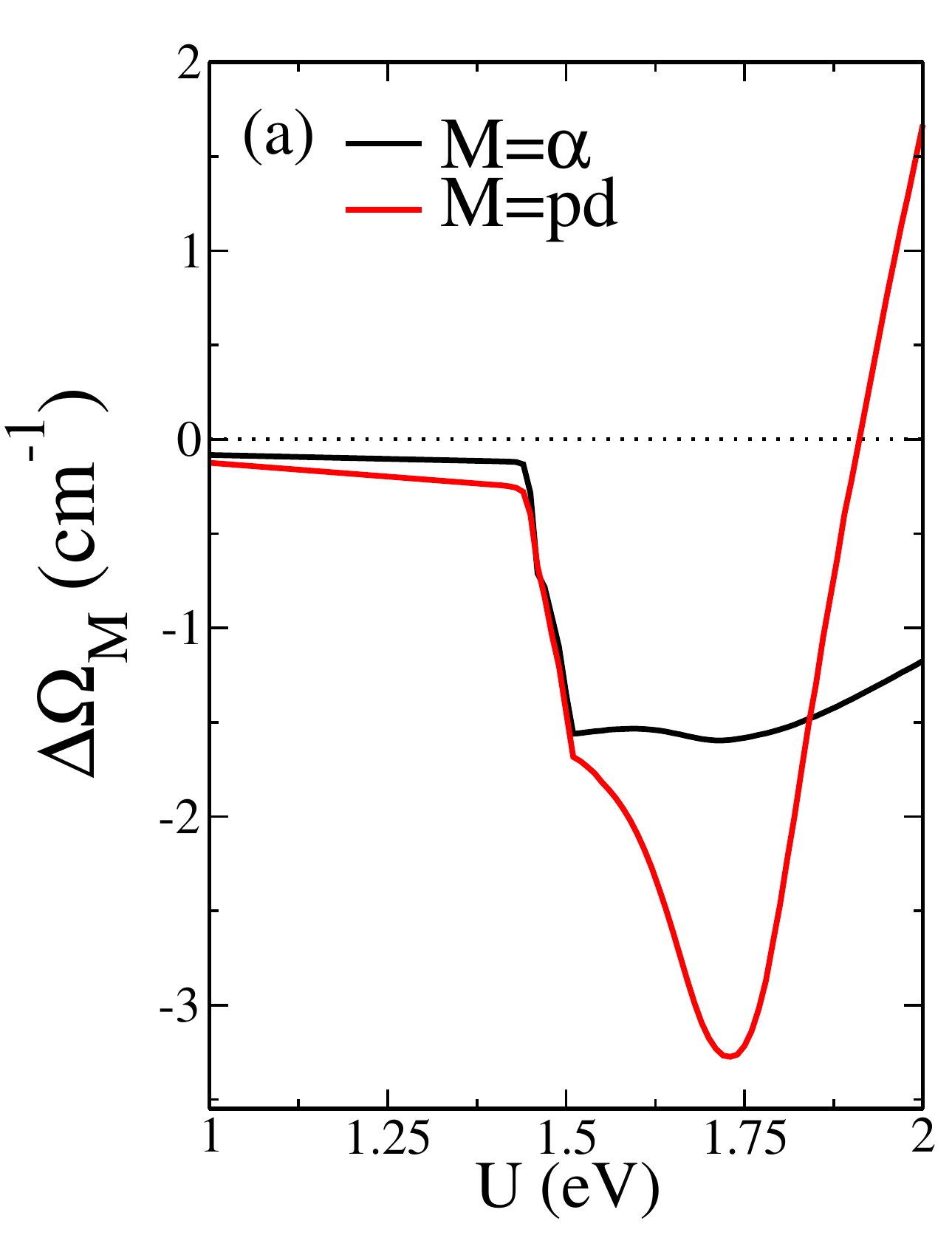}
\includegraphics[clip,width=0.23\textwidth]{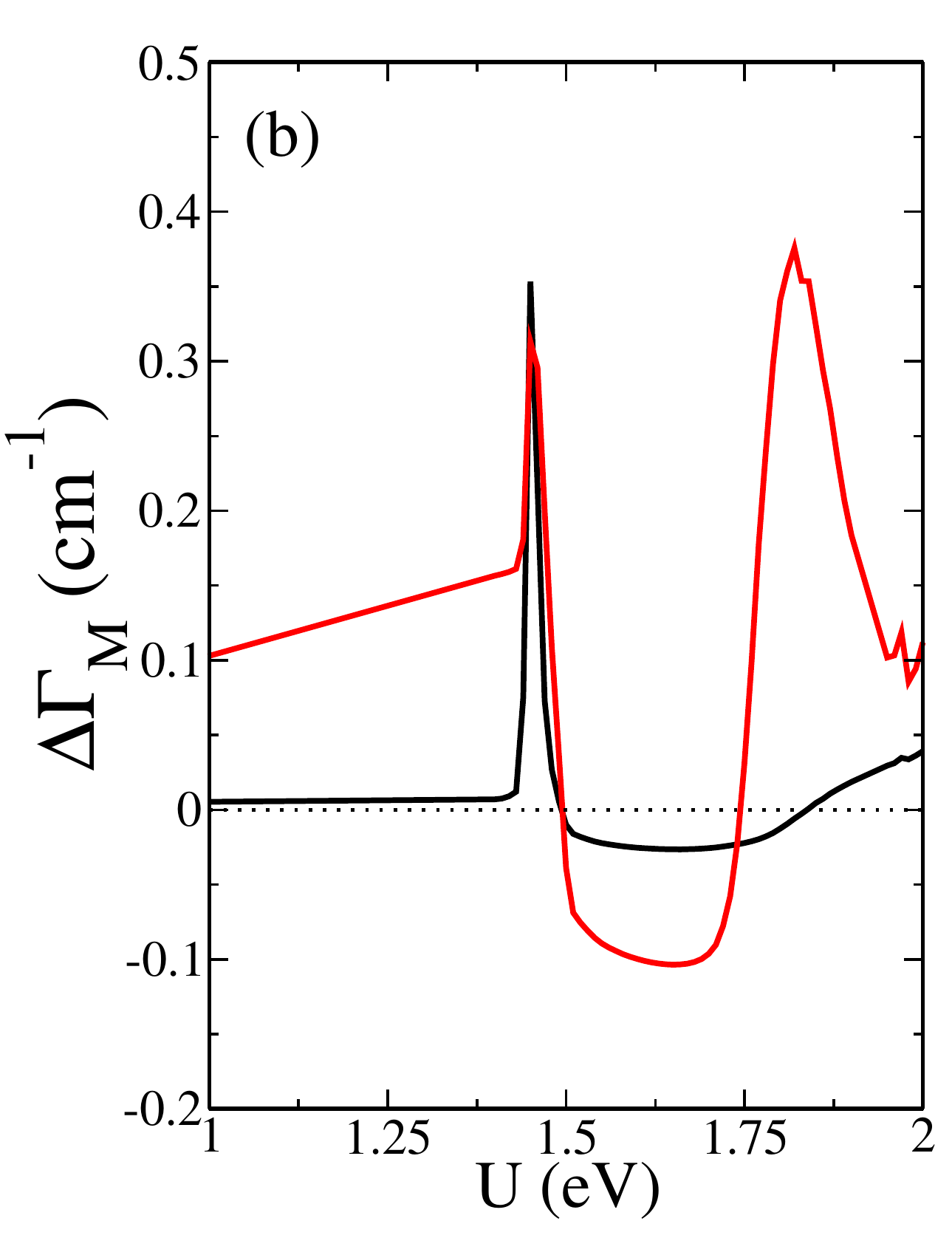}
\caption{ (Color online) Renormalization of the phonon frequency $ \Delta
\Omega_{M}=\Pi'_M(\Omega_{0},U)-\Pi'_M(\Omega_{0}, U=0)$ (a) and phonon
broadening  $\Delta \Gamma_M=-(\Pi''_M(\Omega_{0}, U)-\Pi''_M(\Omega_{0},U=0)$
(b),  for  the electron-phonon couplings  $\hat g^{\alpha}$ and $\hat g^{pd}$.
 $J_H/U=0.25$, $\Omega_0=20$ meV, $\delta h=0.02$ $\rm \AA$ and a
$\delta$-function broadening $\eta=3$ meV  have been used.
}
\label{fig:selfenergy} 
\end{figure}

Fig.~\ref{fig:selfenergy} shows the contributions of the electron-phonon
couplings to the renormalization of the phonon frequency and to the phonon
scattering rate as a function of $U$ (see Section~\ref{sec:phononselfenergy}). To
better visualize the variations of the phonon frequency and phonon broadening 
when entering in the magnetic
state,  we plot $\Delta \Omega_M=\Pi'_M(\Omega_{0},U)-\Pi'_M(\Omega_{0}, U=0)$
in Fig.~\ref{fig:selfenergy} (a) and $\Delta
\Gamma_M=\Gamma_M(U)-\Gamma_M(U=0)=-(\Pi''_M(\Omega_{0},U)-\Pi''_M(\Omega_{0},
U=0))$ in Fig.~\ref{fig:selfenergy} (b).

When entering into the magnetic state ($U \geq 1.45$ eV),  both $\Delta
\Omega_\alpha $ (black) and $\Delta \Omega_{pd}$ (red) are negative, resulting
in phonon softening. This non-intuitive behavior is linked to the multi-band
character of the iron pnictides. Since $\Pi'_M(\Omega)$ is the Kramers-Kronig
integral of $\Pi''_M(\Omega)$, the softening is related to the 
spectral weight redistribution
from high energies ($\Omega>\Omega_{0}$) to
lower energies ($\Omega<\Omega_{0}$) when entering into the magnetic state. In
one band models when a gap opens there is a shift of the spectral weight to
higher energies and hardening is expected. Due to the multiorbital character of
the iron superconductors, the reorganization of the low-energy spectral weight
is non trivial and
part of the spectra shifts closer to the Fermi energy, see for example Fig.~3 in Ref.~[\onlinecite{nosotrasprb13}]. $\Delta \Omega_\alpha$ is
negative in all the range of parameters studied. On the contrary, the
non-monotonic behavior of $\Delta \Omega_{pd}$ results in hardening for $U>
1.9$ eV. The different behavior due to $\hat g^{pd}$ and $\hat g^{\alpha}$ couplings at large
values of $U$ is associated with the different way in which the parameters
$|g^M_{nn'}|^2$ weight the energy excitations around $\Omega_0$ in Eqs.~(\ref{eq:rePi}) and
(\ref{eq:imPi}).

As shown in Fig.~\ref{fig:selfenergy} (b), $\Delta\Gamma_\alpha$
and $\Delta\Gamma_{pd}$ as a function of $U$ are non-monotonic. They change
considerably when entering into the magnetic state. Narrowing (broadening) of
the phonon linewidth corresponds to negative (positive) $\Delta \Gamma_M$. The
large peak at the onset of magnetism at $U=1.45$ eV is due to a particular band
structure reorganization and is not a robust feature for other parameters.  
For larger interactions,
the linewidth shows non-monotonic behavior: with narrowing followed by broadening.

\section{Discussion and comparison to experiments}

Here we discuss our results in comparison with experiment. Note that we address the onset of the magnetic
state as a function of the interaction $U$ at zero temperature, while in an
experiment the varying parameter is the temperature and the interaction $U$
remains constant.
Experimentally, the $A_{1g}$ As-phonon appears in the $A_{1g}$ polarization symmetry in the
paramagnetic state as a small or non-identifiable
peak,~\cite{canfieldprb08,gallaisprb11,sugai12,revramanpnic12} depending
on the compound. This peak is strongly enhanced in 122 compounds when
decreasing the temperature through the magneto-structural
transition.~\cite{canfieldprb08,chauviere09,gallaisprb11,sugai12} 
No phonon peak is observed because of symmetry
in the $B_{1g}$ polarization geometry in the paramagnetic state
whereas a phonon anomaly clearly emerges in the magnetic phase.~\cite{sugai10,gallaisprb11,sugai12,revramanpnic12} In BaFe$_2$As$_2$ the $B_{1g}$ instensity is larger than the one corresponding to the $A_{1g}$ polarization
symmetry.~\cite{gallaisprb11,sugai12}
No peaks are observed in the $B_{2g}$ symmetry in
either state. Our calculations reproduce
the appearance of a peak in the $B_{1g}$ Raman polarization and not in the $B_{2g}$
one in the anisotropic magnetic state without invoking
the structural transition. As discussed in the previous
section, this is a consequence of symmetry and not specific to any particular
electron-phonon coupling. For the same reason no Raman signal is obtained in any of
these symmetries $B_{1g}$ and $B_{2g}$ in the paramagnetic tetragonal state.  
 
Whereas the appearance or not of an $A_{1g}$ As-phonon peak
in the different polarization geometries is quite simple,
being dictated by pure group theory arguments,
the quantitative discussion of the relative intensities is trickier. Each of the couplings here considered, $\hat g^{\alpha}$ and $\hat g^{pd}$, accounts for one of the features observed
experimentally  but none of them alone can explain both.
The coupling via $\hat
g^{\alpha}$ results, for interactions $U>1.8$ eV, in a larger
intensity in the B$_{1g}$ Raman polarization with respect to the
$A_{1g}$ one, as observed experimentally in BaFe$_2$As$_2$.~\cite{gallaisprb11,sugai12} However, this behavior is accompanied by a reduction
of the maximum intensity in the $A_{1g}$  polarization symmetry by an order of
magnitude in the magnetic state.
This is at odds with the strong enhancement of
the $A_{1g}$ peak intensity experimentally observed in 122 compounds.~\cite{canfieldprb08,chauviere09,gallaisprb11,sugai12} An increase of $A_{1g}$ is observed with the coupling $\hat g^{pd}$ but with $I^{A_{1g}}>>I^{B_{1g}}$.

With the estimated couplings, $\hat g^{pd}$ would
dominate the Raman response and $I^{A_{1g}}>I^{B_{1g}}$ would be
expected (see Fig.~\ref{fig:intensities}).
However, as
discussed in Section~\ref{sec:model},  the exact dependence of the
energy integrals on the As position $f(R)$ is not known.  The $1/R^4$ function used is valid
for localized orbitals and could strongly overestimate the electron-phonon
coupling in a covalent system such as the iron pnictides. 
A more realistic functional dependence $f(R)$, with a slower dependence on $R$,
or a difference in the distance dependence of $pd\sigma$ and $pd\pi$ 
could result in a dominance of $\hat g^{\alpha}$ with $I^{B_{1g}}>I^{A_{1g}}$. 
However, note that the relation between magnetization and the system geometry resulting from LDA calculations,~\cite{yndurainprb09} with an increase of the magnetization for elongated tetrahedra, would be consistent with a dominance of the $pd$-dependence on the hoppings, while if they are modified according to the
$\alpha$-dependence, the magnetization decreases (see Fig.~7 in
Ref.~[\onlinecite{nosotrasprb12-2}]).

Experimentally,~\cite{gallaisprb11} the phonon lineshape of undoped
compounds has been found to be strongly
symmetric with a Fano factor $|q|$ bigger than $30$. With electron doping it
acquires an asymmetric shape, with $q \sim -6.5$. The experimental result in
undoped pnictides is in agreement with the values of $q_M$ obtained above in the
orbital differentiation region. With electron doping we expect to enter into the
itinerant region, in which the Fano factor is extremely sensitive to parameters
and no robust prediction can be made.

Both hardening and softening of the $A_{1g}$ As-phonon have been found in experiments when
entering into the magnetic state,~\cite{canfieldprb08,gallais08,rahlenbeckprb09} with changes in the phonon frequency of the
order of $1-3$ cm$^{-1}$ with respect to the zero temperature value. Both electron-phonon coupling and phonon-phonon interaction are expected
to contribute to the frequency renormalization, but it is not obvious how to
separate the two contributions. Our calculations report softening with frequency
renormalizations of the same order of magnitude as experimentally found, except
in the case of $\hat g^{pd}$ coupling at large on-site interactions $U$ which
shows hardening.

Raman experiments have also reported a narrowing of the phonon linewidth which,
depending on the material, ranges from $1$ to $3$ cm$^{-1}$ when
undergoing a magnetic transition.~\cite{canfieldprb08,gallais08,rahlenbeckprb09} For different
on-site interactions, our calculations show both narrowing and broadening,
but with a change in linewidth smaller than observed
experimentally. The largest
values are associated with the electron-phonon coupling channel $\hat g^{pd}$. 
The scattering rate is also very sensitive to the broadening parameter 
$\eta$, related to the electron scattering rate which is reduced in the magnetic
state.

Our results show several features compatible with experimental reports but do not offer a completely satisfactory description of the experiments. It is not clear to us whether the discrepancies arise from the approximations done in the calculations (Hartree-Fock description of interactions and magnetism, neglection of the resonant Raman diagrams, lack of a self-consistent treatment of magnetization and phonons on an equal footing, or the coupling constants estimates) or whether electron-phonon couplings beyond those discussed here should be considered.  Some of these electron-phonon couplings are (a) the  dependence of the Coulombic crystal field $\epsilon^{Coul}_\mu$ on the As position, (b) the dependence of the electronic interaction parameter $U$ on the As-position due to the change in the screening\cite{sawatzky09} or (c) the spin-phonon coupling. \cite{egamiAdvCondMatt10,zhang10}

\section{Summary}

In summary, in this paper we have calculated the Raman spectral properties
of the optical out-of-plane As lattice vibrations (the $A_{1g}$ As-phonon) in the paramagnetic
and in the $(\pi,0)$ magnetic states of the iron pnictides.
Using a tight binding Hamiltonian\cite{nosotrasprb09} based on the Slater-Koster approach,
we have identified two qualitatively different sources
of electron-phonon coupling: one related to the Fe-As-Fe
angle $\alpha$ ($\hat g^{\alpha}$), and one related to the
Fe-As energy integrals $pd\sigma$ and $pd\pi$ ($\hat g^{pd}$).
Both of them contain a  {\it local} ($\bf k$ independent) term
and a {\it non-local} ($\bf k$ dependent) term, associated
with the phonon modulation of the atomic Fe energy levels
and with the effective Fe-Fe hopping amplitudes, respectively.
The magnetic order has been taken into account by means
of a mean-field Hartree-Fock of the electronic Hamiltonian.~\cite{nosotrasprl10,nosotrasprb12-2}
The Raman response of the $A_{1g}$ As-phonon has been calculated
using a suitable generalization of the charge-phonon theory\cite{rice-chphonon-prl76} to the Raman scattering.~\cite{emm,emmprb12}

Our results indicate that a finite Raman intensity can be observed in the magnetic state in the $B_{1g}$ but not in the $B_{2g}$ polarization and it is a consequence of the coupling of the phonons to an anisotropic electronic state with non-equivalent $x$ and $y$ directions. Electron-phonon coupling via $\hat g^{\alpha}$ can result in a Raman signal larger in the B$_{1g}$ symmetry than in the A$_{1g}$ symmetry, as observed experimentally in BaFe$_2$As$_2$. On the other hand, with $\hat g^{\alpha}$ coupling the A$_{1g}$ Raman intensity strongly decreases in the magnetic state, contrary to the experimental results. Coupling via $\hat g^{pd}$ produces the opposite behavior: a very large enhancement of the A$_{1g}$ intensity in the magnetic 
state, which stays much larger than the B$_{1g}$ intensity in all the range of parameters studied. Due to uncertainties in the absolute values of the couplings, it is neither possible to know the intensity resulting from the sum of both  $\hat g^{\alpha}$ and $\hat g^{pd}$ nor to address careful comparison with experiments. 

For most values of the electronic interactions, the electron-phonon coupling induces softening of the phonon frequency in the magnetic state as compared to the paramagnetic state. This behavior is ascribed to the multi-orbital character of the iron superconductors. Hardening is observed for large values of the interaction $U$ when coupling happens via $\hat g^{pd}$. Narrowing or broadening of the phonon line can appear in the magnetic state depending on the parameters.

With symmetry arguments similar to the ones used
above, a finite phonon
intensity in the $B_{1g}$ symmetry would be also expected in a nematic state \cite{kivelson08,schmalianprb12} in
the absence of magnetism.~\cite{footnote-yamase} We also predict that in the double stripe magnetic state of FeTe, with non-equivalent diagonals, the
out-of-plane $A_{1g}$ Te-phonon acquires a finite Raman intensity
in the $B_{2g}$ polarization geometry, but not in the $B_{1g}$
symmetry.
It would be interesting to explore these possibilities experimentally.

We thank Yann Gallais for useful discussions and for sharing unpublished data with us. We have also benefited from conversations with Thomas Frederiksen, Jorge I\~niguez, Lex Kemper, Indranil Paul and F\'elix Yndurain. We acknowledge funding from MINECO-Spain through Grants FIS2008-00124, FIS2009-08744, FIS2011-29689 and FIS2012-33521.
S.C. acknowledges support from Spanish Education Ministry programme SAB2010-0107.
E.C. acknowledges support from the European FP7 Marie
Curie project PIEF-GA-2009-251904  and Italian Project
PRIN ``GRAF'' n. 20105ZZTSE.

\begin{widetext}
\appendix
\section{}
\label{app:A}
The local terms of the electron-phonon couplings $\hat{g}^{\alpha,\rm loc}$ and $\hat{g}^{pd,\rm loc}$ are calculated from the derivatives of the crystal field terms $\epsilon^{ind}_{\mu}$ corresponding to virtual Fe-As forth and back transitions, see Eqs.~(\ref{eq:galpha_loc}) and (\ref{eq:gpd_loc}). The expressions for these terms are calculated to second order in perturbation theory as detailed in Ref.~[\onlinecite{nosotrasprb09}] and are given here

\begin{eqnarray}
\epsilon^{ind}_{xy,xy}&=&{1 \over {|\epsilon_p - \epsilon_d|}} \left[{1 \over 2} \cos^2\alpha \left(4 pd\pi^2(\cos(2\alpha)-1)-3pd\sigma^2(\cos(2\alpha)+1)\right) \right] \, ,\\
\epsilon^{ind}_{yz,yz}&=&\epsilon^{ind}_{zx,zx}={1 \over {|\epsilon_p - \epsilon_d|}} \left[pd\pi^2 (\cos(2\alpha)-\cos(4\alpha)-2)+{3 \over 4}pd\sigma^2(\cos(4\alpha)-1) \right] \, ,\\
\epsilon^{ind}_{3z^2-r^2,3z^2-r^2}&=&{1 \over {|\epsilon_p - \epsilon_d|}} \left[{3 \over 2} pd\pi^2 (\cos(4\alpha)-1)+pd\sigma^2(12 \cos(2\alpha)-9 \cos(4\alpha)-11)/8 \right] \, ,\\
\epsilon^{ind}_{x^2-y^2,x^2-y^2}&=&{1 \over {|\epsilon_p - \epsilon_d|}} \left[-4 pd\pi^2 \cos^2(\alpha)\right]\, .
\end{eqnarray}
$\epsilon_p$ and $\epsilon_d$ are the onsite energies for the As p-orbitals and for the Fe d orbitals. $\alpha$ is the angle formed by the Fe-As bond and the Fe-plane, see Fig.~\ref{fig:red}. $pd\sigma$ and $pd\pi$ are the energy integrals with values $pd\sigma^2/(\epsilon_d-\epsilon_p) \approx 1$ eV and $pd\pi/pd\sigma =-0.5$ respectively. 
\end{widetext}

\bibliography{pnictides}

\begin{thebibliography}{69}
\bibitem[{not()}]{note-symmetry}
\bibinfo{note}{The symmetry of the electron-phonon system determines whether
  $I^\lambda_{M}$ is finite or zero. $B_{1g}$ and $B_{2g}$ symmetries are
  orthogonal to the $A_{1g}$ phonon symmetry in a tetragonal state. As a
  consequence, $I^{B_{1g}}_M$ and $I^{B_{2g}}_M$ vanish in the paramagnetic
  state (with tetragonal symmetry).}
\bibitem[{foo()}]{footnote-yamase}
\bibinfo{note}{A different contribution to the Raman spectrum in a nematic
  state has been studied in Ref.~\onlinecite{yamase13}.}
\end{thebibliography}

\end{document}